\documentclass[useAMS,usenatbib]{mn2e}

\usepackage{epsfig}

\newcommand{\nc}{\newcommand}

\nc{\be}[1]{\begin{equation}\mbox{$\label{#1}$}}
\nc{\bea}[1]{\begin{eqnarray} \mbox{$\label{#1}$}}
\nc{\Section}[2]{\section{#2}\label{#1}}
\nc{\Bibitem}[1]{\bibitem{#1}}
\nc{\Label}[1]{\label{#1}}

\nc{\eea}{\end{eqnarray}}
\nc{\ee}{\end{equation}}

\def\ltsima{$\; \buildrel < \over \sim \;$}
\def\simlt{\lower.5ex\hbox{\ltsima}}
\def\gtsima{$\; \buildrel > \over \sim \;$}
\def\simgt{\lower.5ex\hbox{\gtsima}}

\title
[Cosmic acceleration and extra dimensions: constraints on modifications 
of the Friedmann equation]
{Cosmic acceleration and extra dimensions: constraints on modifications 
of the Friedmann equation}
\author[\O.~Elgar\o y and T.~Multam\"{a}ki]
{\O ystein Elgar\o y$^{1,2}$\thanks{E-mail: oelgaroy@astro.uio.no} and 
Tuomas Multam\"{a}ki$^2$\thanks{E-mail:tuomas@nordita.dk} \\
$^1$Institute of theoretical astrophysics, University of Oslo, Box 1029,
    0315 Oslo, Norway\\
$^2$NORDITA, Blegdamsvej 17, DK-2100 Copenhagen, Denmark\\
}

\date{\today}
\pagerange{\pageref{firstpage}--\pageref{lastpage}}
\pubyear{2004}
\begin{document}

\maketitle

\label{firstpage}

\begin{abstract}
An alternative to dark energy as an explanation for the present phase of 
accelerated expansion of the Universe is that the  
Friedmann equation is modified, e.g. by extra dimensional gravity, 
on large scales. We explore a natural parametrization of a general 
modified Friedmann equation, and find that the present supernova 
type Ia and cosmic microwave background data prefer a
correction of the form $1/H$ to the Friedmann equation over
a cosmological constant. We also explore the 
constraints that can be expected in the future, and find that there 
are good prospects for distinguishing this model from the standard 
cosmological constant to very high significance if one combines supernova 
data with a precise measurement of the matter density.

\end{abstract}

\begin{keywords}
cosmology:theory -- cosmology:observations -- cosmological parameters 
\end{keywords}

\section{Introduction}

There is mounting evidence that we are living in a universe dominated 
by a dark energy component, acting as a source of gravitational repulsion 
causing late-time acceleration of the expansion rate. Early hints 
came from the classical test of using the magnitude-redshift relationship 
with galaxies as standard candles (Solheim 1966), but the reality 
of cosmic acceleration was not taken seriously until the 
magnitude-redshift relationship was measured recently using high-redshift 
supernovae type Ia (SNIa) (Riess et al. 1998; Perlmutter et al. 1999). 
Cosmic acceleration 
requires a contribution to the energy density with negative pressure, 
the simplest possibility being a cosmological constant. Independent 
evidence for a non-standard contribution to the energy budget of 
the universe comes from e.g. the combination of the cosmic microwave 
background (CMB) and the large-scale 
structure (LSS) of the Universe:  
the position of the first peak in the CMB is consistent 
with the universe having zero spatial curvature, which means that 
the energy density is equal to the critical density. Since  
observations of the LSS show that the contribution of standard 
sources of energy density, whether luminous or dark, is only a 
fraction of the critical density, an extra, unknown component 
is needed to account for the spatial flatness of the Universe 
(Efstathiou et al. 2002; Tegmark et al. 2003).  

The simplest explanation for the present accelerated phase of expansion 
is to re-introduce Einstein's cosmological constant, $\Lambda$.  
The resulting model with $\Lambda$, baryons, radiation, and cold dark 
matter (CDM) is consistent with all large-scale cosmological observations 
like the anisotropies in the CMB radiation 
and the power spectrum of galaxies (Tegmark et al. 2003).    
However, the value of $\Lambda$ implied by the observations is tiny  
compared to the value inferred from the fact that it quantifies the 
energy of the vacuum in quantum field theory.  Faced with this problem, 
the popular choice is to set $\Lambda$ to zero and invoke a new component 
to explain the acceleration.  This does not 
solve the problem of the smallness of $\Lambda$; if it is indeed equal 
to zero, one still needs to understand the physical mechanism behind this.  
Nevertheless, one may hope that $\Lambda=0$ 
may be easier to explain than a small, but non-zero $\Lambda$.  
The question is then what the unknown component driving the accelerated 
phase of expansion is.  One needs to introduce a component with negative 
pressure, and this can be done e.g. by invoking a slowly evolving scalar 
field (Wetterich 1988; Peebles \& Ratra 1988; Ratra \& Peebles 1988), or 
a negative-pressure fluid, e.g. a Chaplygin gas (Kamenshchik, Moschella 
\& Pasquier 2001; Bilic, Tupper \& Viollier 2001).   
A negative-pressure fluid, however, can 
be problematic due to its fluctuations. Fluctuations of the unknown
fluid can grow very rapidly and hence LSS surveys can place strict
constraints on such models, see e. g. \cite{bean,sandvik}. 
In order to circumvent this, it is sometimes assumed that the new fluid 
does not fluctuate on the scales of interest and its existence is 
visible only through modified background evolution. 

A different point of view, which we follow here, is that the late time 
acceleration is not due to an unknown component but rather that 
the Friedmann equation is modified on large scales, e.g. due to extra 
dimensional physics. 

The structure of this paper is as follows.  In section 2 we motivate the 
modified Friedmann equation we discuss in later sections.  In section 3 
consider fits to SNIa data, and in section 4 we extend the fits to 
include CMB data from the Wilkinson Microwave Anisotropy Probe (WMAP).  
In section 5 we discuss future prospects for 
constraining the form of the modified Friedmann equation using 
SNIa data, and section 6 contains our conclusions.

\section{Modified Friedmann equation}

We will consider a modification to the Friedmann equation with no
curvature in the spirit of \cite{dvaliturner}, where they consider
\begin{equation}
\left(\frac{H}{H_0}\right)^2 = \Omega_{\rm m}(1+z)^3+(
1-\Omega_{\rm m})\left(\frac{H}{H_0}\right)^\alpha,
\label{eq:modfriedfinal}
\end{equation}
where $\alpha$ is a parameter restricted to be less than $2$ 
from Big Bang Nucleosynthesis (BBN) considerations. One can motivate Friedmann equations
of this form as general parametrizations of 
the leading effect arising from modified gravity theories. As an example,
consider a simple, single extra-dimensional model  
(Dvali, Gabadadze \& Porrati 2000; Deffayet 2001; Deffayet, Dvali \& Gabadadze 2002).
The effective, low-energy action is given by 
\begin{equation}
S = \frac{M_{\rm Pl}^2}{r_{\rm c}}\int d^4 x dy \sqrt{g^{(5)}}{\cal R} 
+\int d^4 x \sqrt{g}(M_{\rm Pl}^2 R + {\cal L}_{\rm SM}), 
\label{eq:effaction}
\end{equation}
where $M_{\rm Pl}^2 = 1/8\pi G$, $G$ is Newton's gravitational constant, 
$g^{(5)}_{AB}$ is the 5-dimensional metric ($A,B=0,1,2,3,4$), $y$ is the 
extra spatial coordinate, ${\cal R}$ is the 5-dimensional Ricci scalar, 
$g$ is the trace of the 4-dimensional metric, 
$R$ is the 4-dimensional Ricci scalar, and ${\cal L}_{\rm SM}$ is the 
Lagrangian of the fields in the Standard Model.  The first term in Eq. 
(\ref{eq:effaction}) is the bulk 5-dimensional Einstein action, and 
the second term is the 4-dimensional Einstein action localized on the 
brane at $y=0$.  The induced metric on the brane is given by 
$g_{\mu \nu}(x) = g_{\mu \nu}^{(5)}(x,y=0)$.  The quantity $r_{\rm c}$ is 
the new parameter of the theory, and is the crossover scale which sets 
the scale for the transition from 4-dimensional to 5-dimensional gravity. 
For the maximally symmetric Friedmann-Robertson-Walker ansatz 
\begin{equation}
ds_5^2 = f(y,H)ds_4^2-dy^2,
\label{eq:frw5d}
\end{equation}
where $ds_4^2$ is the 4-dimensional maximally symmetric metric, and $H$ is 
the 4-dimensional Hubble parameter, one gets a modified Friedmann equation 
on the brane of the form 
\begin{equation}
H^2\pm\frac{H}{r_{\rm c}} = \frac{8\pi G\rho_{\rm m}}{3},
\label{eq:modfried1}
\end{equation}
where $\rho_{\rm m}$ is the matter density on the brane. This
is often called the Friedmann equation of DGP gravity 
(Dvali, Gabadadze \& Porrati 2000).

Inspired by the above example, one can also consider a generalized 
Friedmann equation 
\be{genfried}
f(H)=H_0^2\Omega_{\rm m}(1+z)^3,
\ee
where instead of modifying the matter content we consider 
modifications of gravity by having an arbitrary function $f$. 
Now assume that there is a critical scale, $H_c$, at
which modifications start to have an effect. Such a scale will
be close to the present Hubble parameter.
At early times, when $H\gg H_c$, we know e.g. from nucleosynthesis 
constraints that $f(H)\approx H^2$. In general we can then expand 
$f$ in terms of $H_c/H$: 
\be{genfried2}
H^2\sum_{n}c_n\Big({H_c\over H}\Big)^n=H_0^2\Omega_{\rm m}(1+z)^3.
\ee
As long as $H\gg H_c$, evolution is standard and therefore $c_0=1$
and terms with $n<0$ must vanish. 
Hence
\be{genfried3}
H^2\Big[1+\sum_{n=1}c_n\Big({H_c\over H}\Big)^n\Big]=H_0^2\Omega_{\rm m}(1+z)^3.
\ee
Non-standard effects start to have an effect at late times i.e. when
$H\sim H_c$. 
Expanding the sum gives
\be{genfried4}
H^2\Big[1+c_1{H_c\over H}+c_2\Big({H_c\over H}\Big)^2+...\Big]=H_0^2\Omega_{\rm m}(1+z)^3,
\ee
from which we see that the cosmological constant is the second order 
correction to the Friedmann equation. The first order correction
corresponds to the DGP model.

Generally, the $n$th order correction for a flat universe is
\be{genfried5}
\Big({H\over H_0}\Big)^2=\Omega_{\rm m}(1+z)^3+(1-\Omega_{\rm m})
\Big({H_0\over H}\Big)^{n-2},
\ee
which is of the same form as (\ref{eq:modfriedfinal})
with $\alpha=2-n$. 

In general, one can consider constraints on the different coefficients
$c_n$ with $n=1,2,...$, much like is done in parametrizing
dark energy (Alam et al. 2003).
In this paper, we will for simplicity only consider a single term
that is assumed to be the leading correction. The power of the 
correction is allowed to be arbitrary i.e. we do not restrict it
to discrete values. This
approach gives information on what the leading order term is, and
gives an idea how well one constrain the different 
terms in the expansion with current and future data.

\section{Constraints from Supernovae Type Ia}

The first test any model attempting to explain the accelerated universe 
must pass is, of course, the SNIa data.  We will in the following 
use the sample of 194 SNIa presented in \cite{barris}. 
The parameters we fit to the data are $\Omega_{\rm m}$ and $\alpha$.  
We consider values $0 < \Omega_{\rm m} < 1$ and $-30 < \alpha < 2$.  
The upper limit on $\alpha$ comes from the limits on the amount of 
energy density present at the epoch of BBN  
(Dvali \& Turner 2003).  The Hubble parameter $h$ is also involved in 
the fits, but is of little interest here, and we marginalize over it.  
The fit to the supernova data involves the luminosity distance 
$d_{\rm L} = c(1+z)\int_0^z dz/H(z)$, and we obtain $H(z)$ for given 
$\Omega_{\rm m}$ and $\alpha$ by solving Eq. (\ref{eq:modfriedfinal}) 
with a Newton-Raphson algorithm.  
\begin{figure}
\begin{center}
{\centering
\mbox
{\psfig{figure=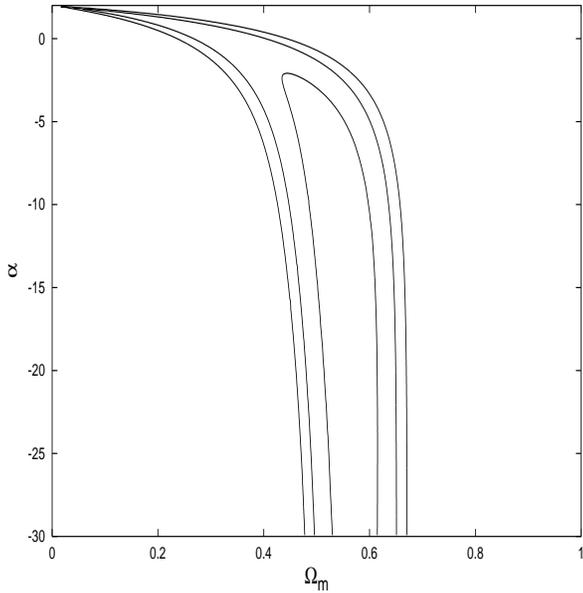,height=8cm,width=8cm}}
}
\caption{68, 95 and 99\% confidence contours for $\Omega_{\rm m}$ and 
$\alpha$, resulting from fitting the collection of supernova data and 
marginalizing over $h$.}
\label{fig:fig3}
\end{center}
\end{figure}
The minimum $\chi^2$ for the model was 195.7 for 191 degrees of freedom, 
with the best-fitting parameters $\Omega_{\rm m}=0.56$, $\alpha=-14.8$.   
The two-parameter confidence contours for $\Omega_{\rm m}$ 
and $\alpha$ are shown in Fig. \ref{fig:fig3}.  It is clear that 
$\alpha$ is very weakly constrained by the supernova data, and can 
seemingly become arbitrarily large and negative. 
We next turn to the CMB data to see if they can provide firmer 
constraints.

\section{Fits to CMB data}

\begin{figure}
\begin{center}
{\centering
\mbox
{\psfig{figure=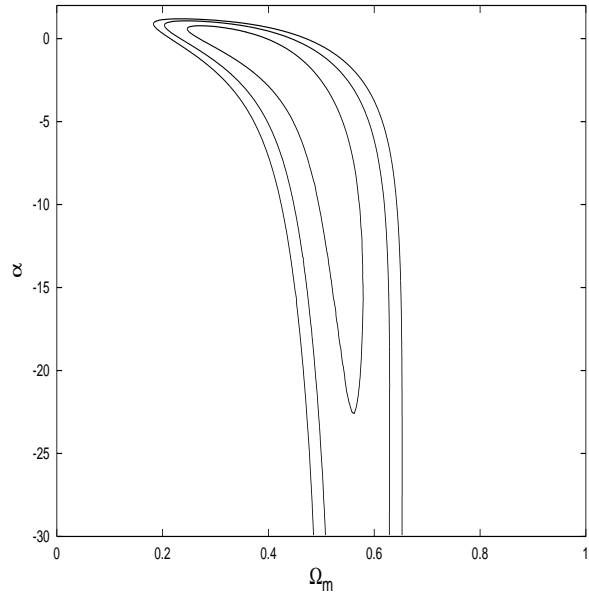,height=8cm,width=8cm}}
}
\caption{68, 95 and 99\% confidence contours for $\Omega_{\rm m}$ and 
$\alpha$, resulting from fitting the collection of supernova data, and 
adding the constraint from the CMB shift parameter.}
\label{fig:fig4}
\end{center}
\end{figure}

Before embarking on a full fit to the CMB data, we first consider the 
much simpler approach of just fitting the so-called CMB shift parameter 
\begin{equation}
{\cal R} = \sqrt{ \Omega_{\rm m}} H_0 r(z_{\rm dec}),
\label{eq:shift}
\end{equation}
where $r(z)=\int_0^z dz/(H(z)/H_0)$ is the comoving distance in a 
flat universe, and 
$z_{\rm dec}$ is the redshift at decoupling.   
The shift parameter describes the shift in the CMB angular power spectrum 
when the 
cosmological parameters are varied 
(Bond, Efstathiou \& Tegmark 1997; Melchiorri et al. 2002; \"{O}dman et 
al. 2003).
From WMAP, $z_{\rm dec}=1088^{+1}_{-2}$, and ${\cal R}_{\rm obs} 
= 1.716\pm 0.062$ (Spergel et al. 2003).   
Adding this constraint to the supernova fit 
is now straightforward (Wang \& Mukherjee 2004).  For each model, we compute $\cal{R}$ from 
Eq. (\ref{eq:shift}) and add $\chi^2_{\cal{R}} = ({\cal R}-{\cal R}_{\rm obs}) 
^2 / \sigma_{\cal{R}}^2$, where $\sigma_{\cal{R}}=0.062$ 
to the $\chi^2$ for the supernova data.  The resulting confidence contours 
in the $\Omega_{\rm m}$--$\alpha$ plane are shown in fig. \ref{fig:fig4}. 
From the figure it is evident that adding the constraint coming from the 
shift parameter leads to a somewhat less degenerate 
range of $\alpha$, especially
when considering the $1\sigma$ contour. Still, even within the
$1\sigma$ limits, $\alpha$ can be as small as $-20$. Such a small
value corresponds to a universe that becomes dominated by the
non-standard terms in the Friedmann equation very quickly after $H<H_0$.

The shift parameter is a useful tool for constraining cosmological
models quickly and very easily using information from the CMB.
However, it does not constrain cosmological 
parameters too well. For example, in the model considered in this paper,
considering only the shift parameter gives a large degeneracy along the
$\alpha$ axis, much like in the SNIa fit. Hence, in order to further 
constrain the parameter plane, one needs to consider the
full shape of the CMB power spectrum.

When fitting the CMB data with a model based on an extra dimensional model,
one should in principle start from the full extra dimensional theory.
This can be problematic due to the bulk-brane interactions and 
hence we will in the following use a simplified approach where we solve 
the standard 4-d perturbed Boltzmann and fluid equations, but with 
the background evolution given by Eq. (\ref{eq:modfriedfinal}).  
For convenience, we will parametrize the effect of the extra term 
in the Friedmann equation by a dark energy component with an 
effective equation of state $w(a)$. As long as the extra component
does not fluctuate, i.e. it only has an effect on the background
evolution, such a parametrization is equivalent to modifying
the Friedmann equation. 

The effective equation of 
state is derived as follows. A dark energy component with equation of state 
$w(a)$ has a density which varies with the scale factor $a$ according to 
\begin{equation}
\rho_{\rm D}(a) = \rho_{0\rm D} \exp\left\{-3\int_1^{a} \frac{da'}{a'}[1+w(a')]\right\}, 
\label{eq:eq2}
\end{equation}
so for a flat universe, $\Omega_{\rm D}=1-\Omega_{\rm m}$, the 
standard Friedmann equation is 
\begin{equation}
\left(\frac{H}{H_0}\right)^2 = {\Omega_{\rm m}\over a^3} 
+(1-\Omega_{\rm m})\exp\left\{-3\int_1^{a} \frac{da'}{a'}[1+w(a')]\right\}.
\label{eq:eq3}
\end{equation}
Comparing this to eq. (\ref{eq:modfriedfinal}), we see that we must have 
\begin{equation}
\exp\left\{-3\int_1^{a} \frac{da'}{a'}[1+w(a')]\right\} = \left(\frac{H}
{H_0}\right)^\alpha,
\label{eq:eq4}
\end{equation}
for the two expressions to match.  By taking the natural logarithm and 
differentiating with respect to $a$ on both sides, we get 
\begin{equation}
-3\frac{1+w(a)}{a}= \alpha\frac{d\ln(H/H_0)}{da},
\label{eq:eq5}
\end{equation}
and since $da = a d\ln a$, we can write this as 
\begin{equation}
w(a) = -1 - \frac{\alpha}{3}\frac{d\ln(H/H_0)}{d\ln a}.
\label{eq:eq6}
\end{equation}
Since $a=(1+z)^{-1}$, we can also write 
\begin{equation}
\frac{d}{d\ln a} = -(1+z)\frac{d}{dz},
\label{eq:eq7}
\end{equation}
giving 
\begin{equation}
w(z) = -1 + \frac{\alpha}{3}\frac{(1+z)}{H/H_0}\frac{d(H/H_0)}{dz}.
\label{eq:eq8}
\end{equation}
Going back to eq. (\ref{eq:modfriedfinal}), 
we can differentiate with respect to 
$z$ and get 
\begin{eqnarray}
2\left(\frac{H}{H_0}\right)\frac{d(H/H_0)}{dz} &= &
\alpha(1-\Omega_{\rm m})\left(\frac{H}{H_0}\right)^{\alpha - 1}
\frac{d(H/H_0)}{dz} \nonumber \\ 
&+& 3\Omega_{\rm m}(1+z)^2,
\label{eq:eq9}
\end{eqnarray}
so that 
\begin{equation}
\frac{d}{dz}\left(\frac{H}{H_0}\right) = \frac{3\Omega_{\rm m}(1+z)^2}
{2\left(\frac{H}{H_0}\right)-\alpha (1-\Omega_{\rm m})\left(\frac{H}{H_0}
\right)^{\alpha - 1}}.
\label{eq:eq10}
\end{equation}
This saves us the trouble of taking numerical derivatives in practical work, 
since $w(z)$ now can by expressed in terms of $H(z)/H_0$ as 
\begin{equation}
w(z)=-1 + \frac{\alpha\Omega_{\rm m}(1+z)^3}{2\left(\frac{H}{H_0}\right)^2 
-\alpha(1-\Omega_{\rm m})\left(\frac{H}{H_0}\right)^\alpha}.
\label{eq:eq11}
\end{equation}
So, instead of modifying the Friedmann equation directly, we can consider a 
standard Friedmann equation, Eq. (\ref{eq:eq3}), 
with a fluid whose equation of state is given by (\ref{eq:eq11}).  
Note that this is an exact reformulation of the background evolution.  
We have checked that fitting the supernova data with this reformulation results in confidence contours in agreement with those in Fig. \ref{fig:fig3}. 

For fitting the CMB TT power spectrum 
(Hinshaw et al. 2003; Kogut et al. 2003) we use
the likelihood code provided by the WMAP team\footnote{http:// lambda.gsfc.nasa.gov/} (Verde et al. 2003). 
The CMB power spectra is computed by using CMBFAST code 
(Seljak \& Zaldarriaga 1996),  
version 4.5.1\footnote{http://www.cmbfast.org/}.
For each point in the parameter space $(\Omega_{\rm m},\alpha,h)$,
we calculate the CMB TT power spectrum keeping the amplitude
of the fluctuations a free parameter by finding
the best-fitting amplitude for each set of parameters. In other words,
we only fit the shape and not the amplitude of the power
spectrum. In calculating the CMB power spectrum we use
$\Omega_{\rm b}=0.044$ (so that we vary the density of cold 
dark matter, $\Omega_{\rm c}$) 
and ignore reionization effects.
Parameters $\Omega_{\rm m}\in[0,1]$ and $\alpha\in[1.5,-10]$ 
have uniform priors
and are chosen to cover the most interesting range of parameters.
For the Hubble parameter $h$, we have explored different 
priors: a uniform prior $h\in[0.5,1.0]$ and a Gaussian prior based on the
HST Key Project value $h=0.72\pm 0.08$ (Freedman et al. 2001). 
We marginalize over
$h$ in making all the plots. The choice of prior makes little
difference after marginalization since most of the weight
comes from $h\sim  0.72$. Here we show results with a Gaussian prior
but confidence contours with a uniform prior are essentially
identical.

\begin{figure}
\begin{center}
{\centering
\mbox
{\psfig{figure=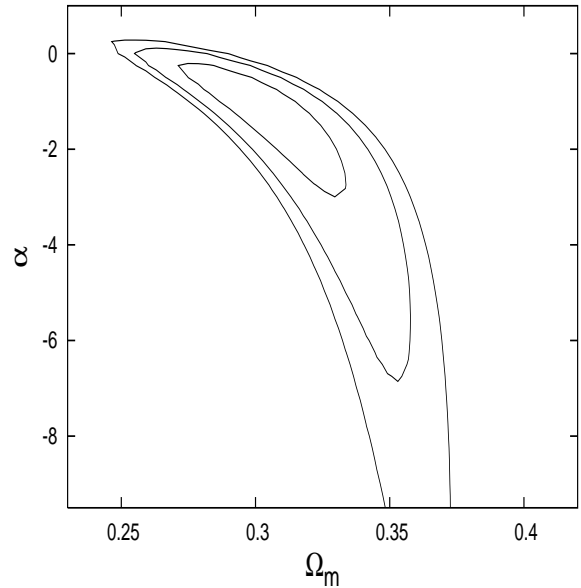,height=8cm,width=8cm}}
}
\caption{CMB constraints (68, 95, and 99 \%) from the WMAP TT power spectrum}
\label{fig:fig5}
\end{center}
\end{figure}
The WMAP TT power spectrum constraints are shown in fig. \ref{fig:fig5}.
From the figure we see that, as expected, having more information 
from the power spectrum than just the shift parameter helps
tighten the constraints significantly. The minimum value of $\alpha$ within the
$1\sigma$ contour is now only about $-3$. Comparing to fig. \ref{fig:fig3},
where only SNIa are used, it is obvious that CMB provides a much
tighter constraint to the model in question than just the SNIa results.
This demonstrates how if one restricts oneself to a single 
cosmological probe in studying models with non-standard background 
evolution, the CMB can be a good first choice instead of supernovae.
If the non-standard model involves new cosmological fluctuating fluids, 
then a simple check is provided by 
considering LSS observations which can be effective in constraining fluids 
with a non-zero sound speed.

\begin{figure}
\begin{center}
{\centering
\mbox
{\psfig{figure=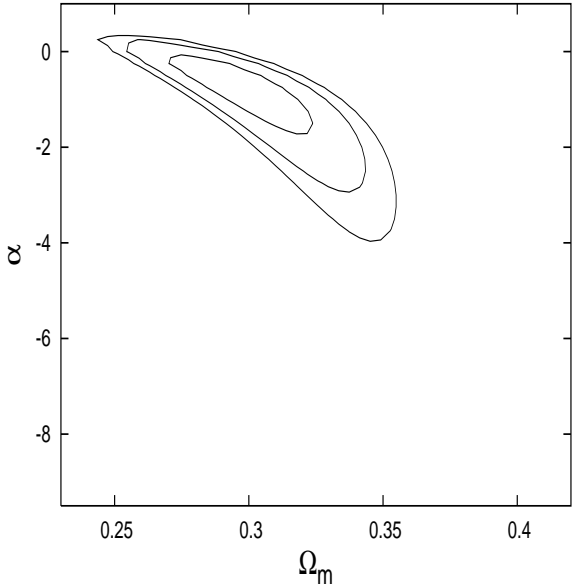,height=8cm,width=8cm}}
}
\caption{Constraints (68, 95, and 99 \%) from combining the 
WMAP TT power spectrum with supernova Type Ia data.}
\label{fig:fig6}
\end{center}
\end{figure}

The combined fit CMB+SNIa is shown in fig. \ref{fig:fig6}.
Adding the SNIa data further relieves the degeneracy along the $\alpha$
axis due to the fact even though both fig. \ref{fig:fig3} and 
\ref{fig:fig5} are both somewhat degenerate along $\alpha$, the
region of degeneracy correspond to different values of $\Omega_{\rm m}$.


\section{Future supernova data}

We have found that the present supernova data cannot put significant 
constraint on the parameter $\alpha$, and that at the present time 
the most stringent constraints comes from the combination of CMB data 
with supernovae.  An interesting question is how well one can do 
with future SNIa surveys.  The planned Dark Energy Probe
\footnote{http://universe.gsfc.nasa.gov/program/darkenergy.html}/ 
Supernova Acceleration Probe\footnote{http://snap.lbl.gov/} is expected 
to observe about 2000 
supernovae type Ia per year out to a redshift of $z\sim 1.7$ 
(Aldering et al. 2002), and this 
should improve the power of this probe to constrain dark energy models 
considerably.  We will in the following simulate data sets of this type, 
following the approach of Saini, Weller \& Bridle (2004),   
and consider how well they can constrain the the model under investigation 
in this paper.  

Empirically, SNIa are very good standard 
candles with a small dispersion in apparent magnitude $\sigma_{\rm mag}=0.15$, 
and there is no indication of redshift evolution.  The apparent magnitude 
is related to the luminosity distance through 
\begin{equation}
m(z) = {{\cal M}} + 5\log D_{\rm L}(z),
\label{eq:appmag}
\end{equation}
where ${{\cal M}}=M_0 + 5\log[(c/H_0)\,{\rm Mpc}^{-1}]+25$.  The quantity 
$M_0$ is the absolute magnitude of type Ia supernovae, and $D_{\rm L}(z) 
= H_0d_{\rm L}(c)/c$ is the Hubble constant free luminosity distance.  
The combination of the absolute magnitude and the Hubble constant, ${{\cal M}}$, can be calibrated by low redshift supernovae 
(Hamuy et al. 1993; Perlmutter et al. 1999). 
The dispersion in the magnitude, $\sigma_{\rm mag}$, is related to the 
deviation in the distance, $\sigma$, by 
\begin{equation}
\frac{\sigma}{d_{\rm L}(z)} = \frac{\ln 10}{5}\sigma_{\rm mag}. 
\label{eq:errdL}
\end{equation}
In our simulated data sets, we assume that the errors in the luminosity 
distance are Gaussian and given by Eq. (\ref{eq:errdL}).  We neglect 
systematic errors.  Furthermore, we assume that the supernovae type Ia are 
uniformly distributed and bin them in 50 redshift bins, giving a relative 
error in the luminosity distance in each bin of $\sim 1\%$.  
We do not add noise to the simulated $d_{\rm L}$, and hence our results  
give the ensemble average of the parameters we fit to the simulated 
data sets.  

\begin{figure}
\begin{center}
{\centering
\mbox
{\psfig{figure=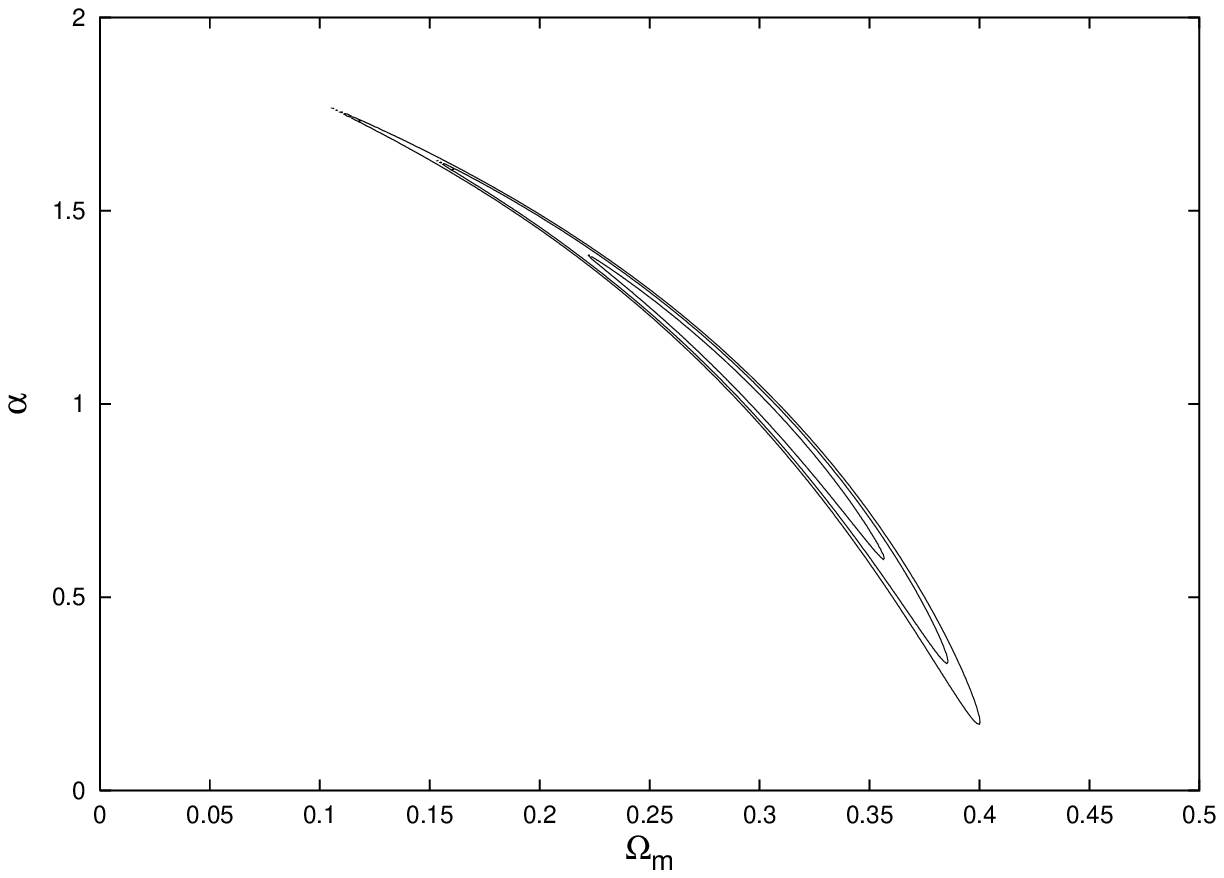,height=8cm,width=8cm}}
}
\caption{Constraints on $\Omega_{\rm m}$ and $\alpha$ from simulated 
data based on $\Omega_{\rm m}=0.3$, $\alpha=1$.}
\label{fig:fig7}
\end{center}
\end{figure}

\begin{figure}
\begin{center}
{\centering
\mbox
{\psfig{figure=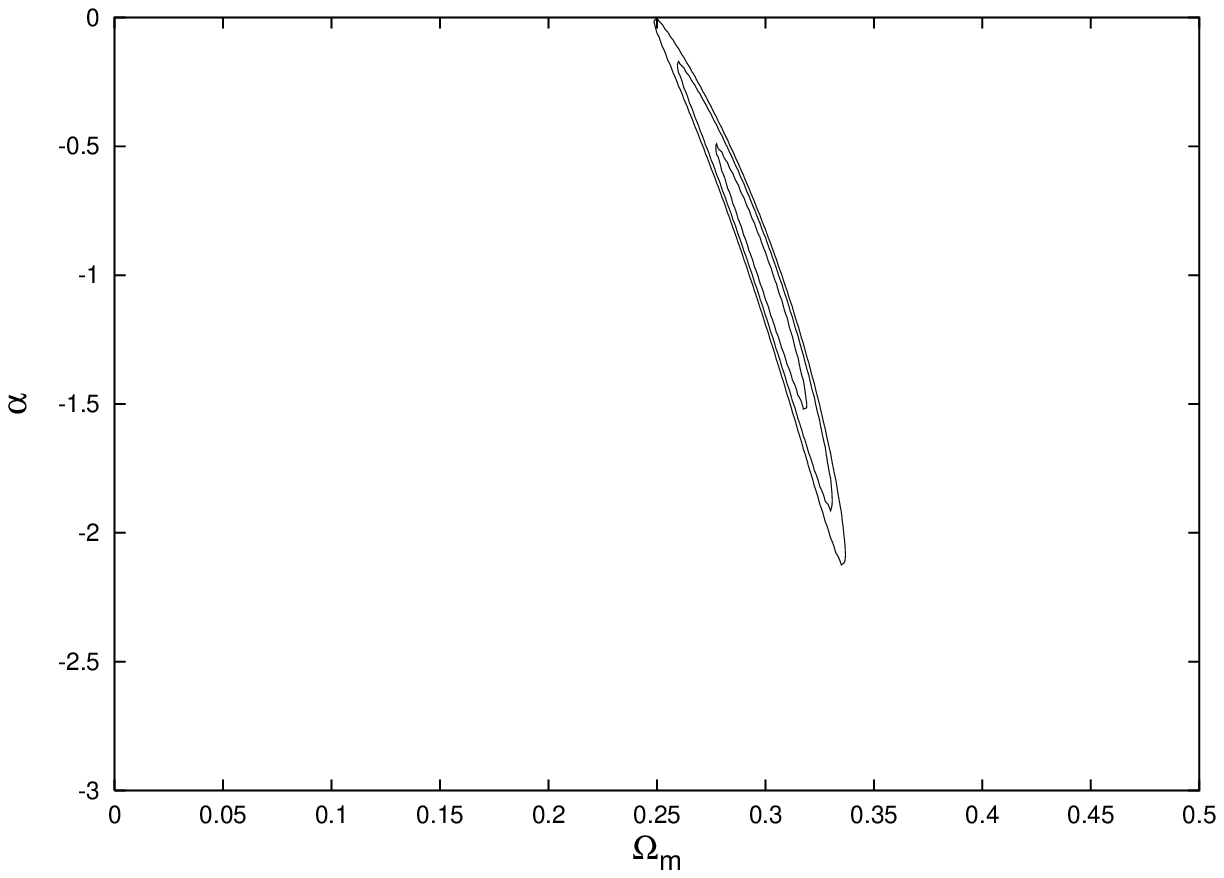,height=8cm,width=8cm}}
}
\caption{Constraints on $\Omega_{\rm m}$ and $\alpha$ from simulated 
data based on $\Omega_{\rm m}=0.3$, $\alpha=-1$.}
\label{fig:fig8}
\end{center}
\end{figure}

\begin{figure}
\begin{center}
{\centering
\mbox
{\psfig{figure=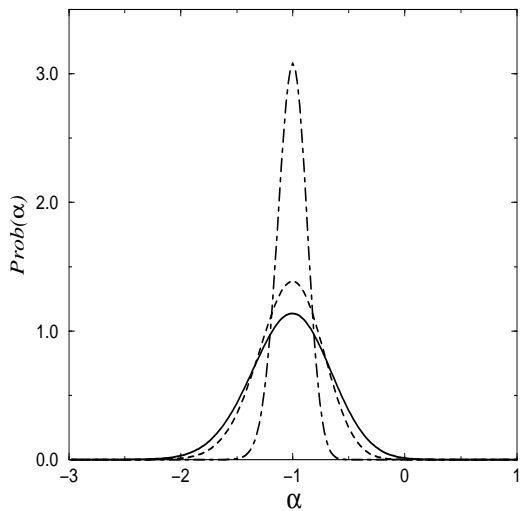,height=8cm,width=8cm}}
}
\caption{Marginalized probability distribution for $\alpha$ with 
three different choices of priors on $\Omega_{\rm m}$: uniform 
$0 < \Omega_{\rm m} < 0.5$ (solid line), Gaussian $\Omega_{\rm m} 
= 0.30\pm 0.02$ (dashed line), and Gaussian $\Omega_{\rm m}= 0.300\pm 
0.005$ (dot-dashed line).}
\label{fig:fig9}
\end{center}
\end{figure}

First, we simulate a data set based on a model with $\Omega_{\rm m}=0.3$, 
$\alpha=1$.  In Fig. \ref{fig:fig7} we show the confidence contours 
resulting from fitting $\Omega_{\rm m}$ and $\alpha$ to this simulated 
data set.  
The constraints which can be derived from a data set of this quality are 
seen to be considerably tighter than those derived from the presently 
available data.  However, there is still a notable degeneracy 
between $\Omega_{\rm m}$ and $\alpha$, so that without a tight prior 
on $\Omega_{\rm m}$, one cannot distinguish between, e.g.,  $\alpha=1$ and 
$\alpha=0.5$.  A similar situation occurs for simulated data based 
on $\Omega_{\rm m}=0.3$, $\alpha=-1$, shown in Fig. \ref{fig:fig8}.  
For this case, we also show in Fig. \ref{fig:fig9} the marginalized, 
normalized probability 
distributions for $\alpha$ for three different priors on $\Omega_{\rm m}$: 
a uniform prior $0 < \Omega_{\rm m} < 0.5$, a Gaussian prior 
$\Omega_{\rm m}  = 0.30\pm 0.02$, and a Gaussian prior $\Omega_{\rm m} 
= 0.300 \pm 0.005$.  Note that only with the last choice of prior, where 
$\Omega_{\rm m}$ is assumed to be known to within $1.7$ \%, 
does one get a really tight constraint on $\alpha$, but that even in 
the case of a uniform prior one can exclude a cosmological constant 
(which corresponds to $\alpha=0$) at 99.9 \% confidence. 

\section{Conclusions}
\label{conclusions}

In this work we have studied observational constraints on
a modified Friedmann equation that mimics dark energy in the
universe. Modifications of the type we have considered here 
may occur naturally in models with large extra dimensions, and 
provide an attractive alternative to introducing an unknown component 
with negative pressure.  We have found that the combination of 
the magnitude-redshift relationship derived from supernovae type Ia 
and the WMAP TT power spectrum constrain the exponent of the extra term in the 
modified Friedmann equation, $(H/H_0)^\alpha$, to be around $-1$.  
From the point of the expanded Friedmann equation, this corresponds
to $n=3$, which is the next term after the cosmological constant
in the expansion (\ref{genfried3}). In particular, a cosmological 
constant, corresponding to $\alpha=0$ or $n=2$ seems to be disfavoured
for the studied parameter space. Furthermore,
the first order correction $n=1$ corresponding to the DGP model
appears to be strongly disfavored over the $n=3$ term.


We have also used simulated data sets of the type one can expect 
from future satellite-based supernova surveys to estimate the accuracy 
with which one can hope to constrain the corrections
to the Friedmann equation. Tight constraints on $\alpha$ can be expected if the 
matter density parameter $\Omega_{\rm m}$ is known accurately from other 
observations.  But even without any priors on $\Omega_{\rm m}$, 
one can rule out a cosmological constant at high significance if the 
true universe is described by $\alpha=-1$.

An obvious omission in this work is that we have not considered 
constraints coming from LSS. Since there is no extra negative 
pressure fluid in the model, one does not expect large deviations 
in the matter power spectrum. Furthermore, the background
evolution follows the standard behaviour until very recently, which
suggests that linear growth will be standard for the most
of the expansion history (Lue, Scoccimarro \& Starkman 2004; Multam\"{a}ki, Gazta\~naga \& Manera 2003). A more detailed analysis
of linear and non-linear growth is left for future work.

Finally, we note that the term we have considered is only the 
leading order correction to the standard Friedmann equation.   
The present data indicate that the first order correction 
is $(H_0/H)$ but other terms can also play a role.
It is an interesting question whether a combination 
of data sets of the quality we can expect in the future can constrain 
the number of correction terms and their form.

\section*{Acknowledgments}
\O E gratefully acknowledges the hospitality of NORDITA, where 
parts of this work were carried out.  
TM is grateful to the Institute of theoretical 
astrophysics, University of Oslo, for their hospitality during
the first stages of this work.

\label{lastpage}

\end{document}